%% file: main.tex
\documentclass[runningheads]{llncs}
\usepackage[T1]{fontenc}
\usepackage{graphicx}
\usepackage{amsmath}
\usepackage{xcolor}
\usepackage{makecell}
\usepackage{multirow}
\usepackage{graphicx}

\begin{document}

\makeatletter
% latex.ltx, line 8262:
\def\@fnsymbol#1{%
   \ifcase#1\or
   \TextOrMath \textdagger \dagger\or
   \TextOrMath\textasteriskcentered *\or
   \TextOrMath \textdaggerdbl \ddagger \or
   \TextOrMath \textsection  \mathsection\or
   \TextOrMath \textparagraph \mathparagraph\or
   \TextOrMath \textbardbl \|\or
   %\TextOrMath {\textasteriskcentered\textasteriskcentered}{**}\or
   \TextOrMath {\textdagger\textdagger}{\dagger\dagger}\or
   \TextOrMath {\textdaggerdbl\textdaggerdbl}{\ddagger\ddagger}\else
   \@ctrerr \fi
}
\makeatother

\makeatletter
\newcommand{\printfnsymbol}[1]{%
  \textsuperscript{\@fnsymbol{#1}}%
}
\makeatother

\title{Multi-modal Contrastive Learning for Tumor-specific Missing Modality Synthesis}
\titlerunning{Multi-modal Contrastive Learning for Missing Modality Synthesis}

\author{
Minjoo Lim\thanks{equal contribution} \and
Bogyeong Kang\printfnsymbol{1} \and
Tae-Eui Kam\thanks{Corresponding author}}

\authorrunning{ M. Lim, B. Kang, et al.}
\institute{Department of Artificial Intelligence, Korea University, Seoul, South Korea.
\email{\{imj700,kangbk,kamte\}@korea.ac.kr}\\
}
\maketitle              % typeset the header of the contribution
%
% ---- Abstract (no citations to be present) ----
%
\begin{abstract}
Multi-modal magnetic resonance imaging (MRI) is essential for providing complementary information about brain anatomy and pathology, leading to more accurate diagnoses. However, obtaining high-quality multi-modal MRI in a clinical setting is difficult due to factors such as time constraints, high costs, and patient movement artifacts. To overcome this difficulty, there is increasing interest in developing generative models that can synthesize missing target modality images from the available source ones. Therefore, our team, PLAVE, design a generative model for missing MRI that integrates multi-modal contrastive learning with a focus on critical tumor regions. Specifically, we integrate multi-modal contrastive learning, tailored for multiple source modalities, and enhance its effectiveness by selecting features based on entropy during the contrastive learning process. Additionally, our network not only generates the missing target modality images but also predicts segmentation outputs, simultaneously. This approach improves the generator's capability to precisely generate tumor regions, ultimately improving performance in downstream segmentation tasks. By leveraging a combination of contrastive, segmentation, and additional self-representation losses, our model effectively reflects target-specific information and generate high-quality target images. Consequently, our results in the Brain MR Image Synthesis challenge demonstrate that the proposed model excelled in generating the missing modality.

% The abstract should briefly summarize the contents of the paper in

\keywords{Brain MRI synthesis  \and Missing modality issue \and Multi-modal learning} 

\end{abstract}
%
% ---- Chapters  ----
%
\input{Chapter/1.Introduction}

\input{Chapter/2.Methods}

\input{Chapter/3.Experiments}

\input{Chapter/4.Conclusion}

%
% ---- Acknowledgements(Optional), Disclosure of Interests ----
%
\begin{credits}
\subsubsection{\ackname} This work was supported by grants from the Institute of Information \& Communications Technology Planning \& Evaluation (IITP)—specifically, the Artificial Intelligence Graduate School Program at Korea University (No. RS-2019-II190079). Additionally, this research received funding from the National Research Foundation of Korea (NRF), supported by the Korea government (MSIT) under grant (No. RS202300212498).

\end{credits}
%
% ---- References (Bibliography) ----
%

\bibliographystyle{splncs04}
\bibliography{ref}

\end{document}

%% file: Chapter/1.Introduction.tex
\section{Introduction}
Reliable brain tumor segmentation in magnetic resonance imaging (MRI) plays a crucial role in diagnosis and treatment \cite{cao2020auto,li2023BraSyn}. 
Since performing this task manually is tedious and highly variable, the various advantages of deep learning-based automated tumor segmentation methods have been studied recently \cite{li2023BraSyn}. 
Most of these recent methods utilize all four commonly used MRI modalities in clinical scenarios, including T1-weighted (T1), contrast enhanced T1-weighted (T1-ce) image, T2-weighted (T2),and Fluid Attenuated Inversion Recovery (FLAIR) image \cite{azad2022review,cao2020auto,li2023BraSyn,wang2023miccai}. 
These various multi-modal MRIs provide complementary information about brain anatomy and pathology, leading to the most accurate diagnoses and effective treatments \cite{azad2022review,meng2024diff}.
\textcolor{black}{However, in clinical practice, MRI sequences are often missing due to time constraints, patient movement artifacts, costs, or high radiation doses \cite{li2023BraSyn,meng2024diff}.} 
To address this issue, there has been growing interest in developing generative models that can synthesize missing modalities from the available ones \cite{li2023BraSyn,meng2024diff}. 
\textcolor{black}{This necessity is also highlighted by the Brain MR Image Synthesis for Tumor Segmentation (BraSyn) challenge, one of the tasks of the Brain Tumor Segmentation (BraTS) challenge 2024 \cite{li2023BraSyn}, in which our team, PLAVE, participated.}

\textcolor{black}{Several studies~\cite{cao2020auto,cao2023AE,cao2023ACA,baltruschat2024pix2} have employed generative adversarial network (GAN) -based models to generate missing target modality from multi-source modalities. Specifically, Cao \textit{et al.}~\cite{cao2020auto} proposed Auto-GAN, which introduces an autoencoder network with a novel self-supervision constraint to provide target-specific information to guide the generator training. An extended version, i.e., AE-GAN, replaces the traditional discriminator with an autoencoder-based discriminator \cite{cao2023AE}.
Another version, i.e., ACA-GAN, effectively extracts complementary information from available modalities by deploying both single-/multi-modal attention modules \cite{cao2023ACA}. These previous studies have shown that GAN-based generative models, self-supervision constraints, and single/multi-modal attention modules can significantly enhance the performance of generating missing modalities.}

\textcolor{black}{Therefore, by extending ACA-GAN~\cite{cao2023ACA}, we design a missing MRI generative model that utilizes contrastive learning with integrated multi-source modalities, with a particular focus on accurately generating tumor regions within the target images.}
Firstly, we integrate multi-modal contrastive learning to ensure that the generator remains invariant to multi-source style variations and learns to focus on the structural information of multi-source modalities. The contrastive learning in natural image translation task is generally used to enhance the learning of structural information within a single source domain \cite{hu2022qs,park2020contrastive}. Meanwhile, since our task deals with multi-modal MRIs, we conduct the multi-modal contrastive learning based on multi-source modalities. We also employ the QS-Attn module~\cite{hu2022qs} to mitigate the issue where limited information in features used for contrastive learning can negatively impact the learning process~\cite{hu2022qs}. This module selects features based on entropy, ensuring that only the most informative features are utilized during our multi-modal contrastive learning. Secondly, our multi-modal translation network not only generates the missing target modality images but also predicts the segmentation output simultaneously.
By incorporating segmentation masks into the training process through segmentation decoder, the network can learn the distinctive features of tumors more effectively from multi-source images.
This approach eventually improves the generator's ability to accurately produce tumor regions, leading to better performance in downstream segmentation tasks.
Additionally, our models are effectively trained by combining contrastive loss, segmentation loss, generator loss, and multiple self-representation losses. To ensure that target-specific information is effectively conveyed to the generator at all stages, we use self-representation losses for decoder features as well as single-modal encoder features and multi-modal fusion features~\cite{cao2023ACA}.
Consequently, through the BraSyn challenge, we demonstrate the performance of our \textcolor{black}{model} by successfully generating high-quality missing modalities from available multi-modal data.

%% file: Chapter/2.Methods.tex
\section{Methods}
\begin{figure*}[t]
\begin{center}
\includegraphics[width=\linewidth]{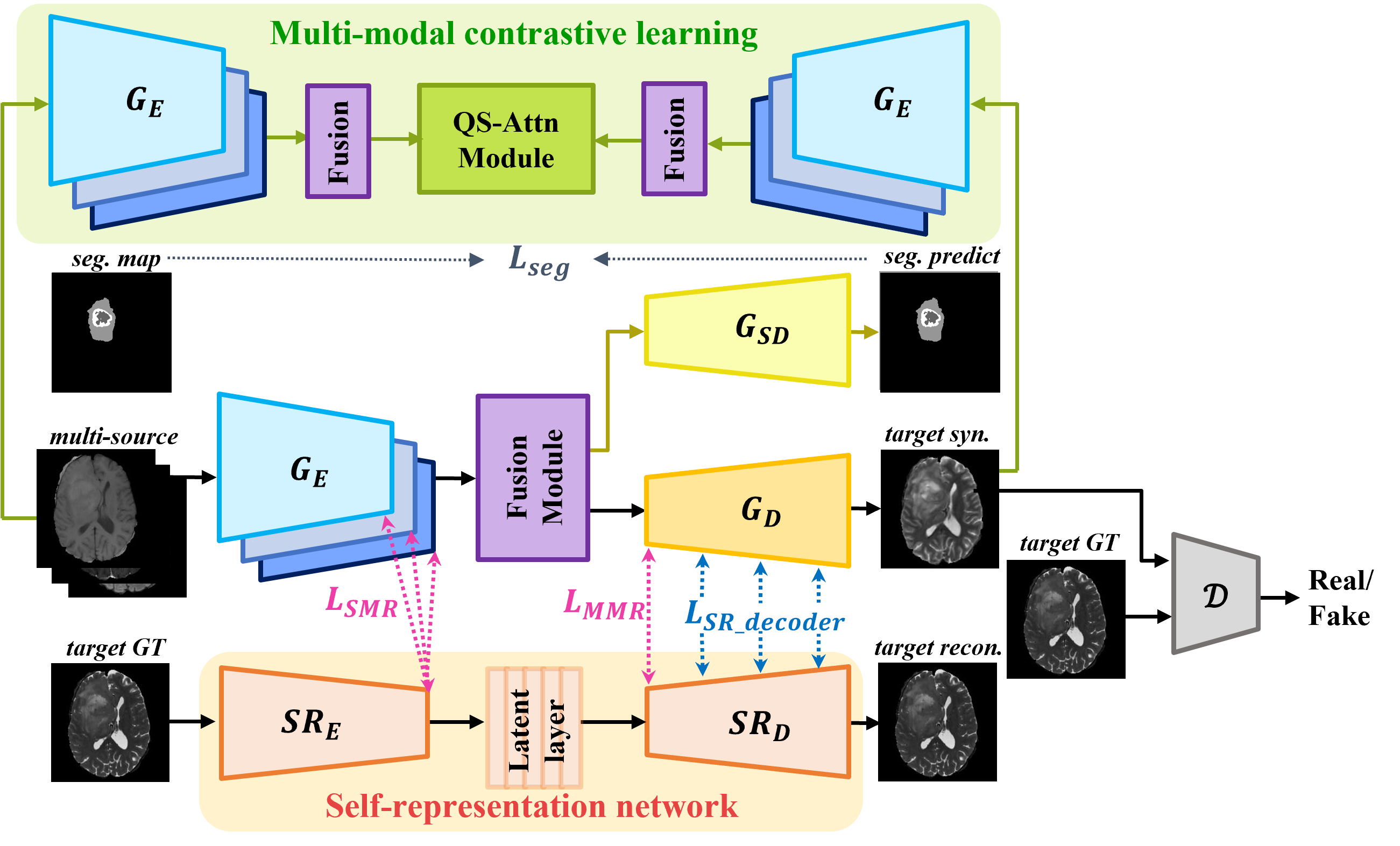}
\end{center}
\caption{Overall framework of our missing MRI generative model.} 
\label{fig:overview}
\end{figure*}

The goal of missing modality synthesis is to generate images for a missing target modality by utilizing multiple available source images. To achieve this, \textcolor{black}{our model is designed} by incorporating multi-modal contrastive learning and a segmentation decoder, enhancing the ability to learn from multi-source modalities. Our \textcolor{black}{model} builds upon the multi-modal translation network $T$ and self-representation autoencoder network $SR$ of the previously studied ACA-GAN~\cite{cao2023ACA}. Specifically, the multi-modal translation network $T$ comprises a generator $G$ and a discriminator $\mathcal{D}$. 
\textcolor{black}{The $G$ comprises a multi-branch encoder ${G}_{E}$  with an attached single attention module, multi-modal fusion module, a decoder ${G}_{D}$ for generating missing modality, and an additional segmentation decoder ${G}_{SD}$.}
\textcolor{black}{The overall framework of our model is shown in Fig.\ref{fig:overview}.}

\subsection{Multi-modal contrastive learning}
\textcolor{black}{Unlike conventional contrastive learning methods \cite{park2020contrastive,hu2022qs}, focusing on single-source modality, our task requires a distinct approach for multi-source modalities.}
\textcolor{black}{To address this, we integrate multi-modal contrastive learning into our multi-modal translation network.}
\textcolor{black}{By utilizing fused features from all modalities, this approach enables contrastive learning on more robust and generalized features that incorporate both intra- and inter-modal complementary information.}

\textcolor{black}{When the source images are $X=\{x_1, x_2, x_3\}$, and the missing target real image is $y$, each of $x_i$ (where $i\in{1,2,3}$) is processed through its respective multi-branch encoder designed for its specific modality.
These modality-specific features are then processed through the multi-modal fusion module to produce the final fused features $f^{X}_{fusion}$.}
\textcolor{black}{The generated image $G(X)$ follows the same processing steps, resulting in its corresponding feature $f^{G(X)}_{fusion}$. We then compute the multi-modal contrastive loss between $f^{X}_{fusion}$ and $f^{G(X)}_{fusion}$.}
The anchor feature $q$ is extracted from the $f^{G(X)}_{fusion}$, serving as the query. The positive $k^+$ is obtained from the feature at the same location in $f^{X}_{fusion}$, and the negative features ${k}^{-}$ are selected from the remaining locations.
The contrastive loss is defined as follows, ensuring that the query is close to the positive $k^+$ while being pushed away from the negative $k^-$~\cite{park2020contrastive}:

\begin{equation}
    {L}_{con}=-\log\Biggl[\cfrac{\exp(q\cdot{k}^{+}/\tau)}{\exp(q\cdot{k}^{+}/\tau)+\sum\nolimits_{n=1}^{N-1}\exp(q\cdot{k}^{-}_{n}/\tau)}\Biggr], \label{eq8}
\end{equation}
\textcolor{black}{where ${k}^{-}_{n}$ denotes $n$-th negative, and $\tau$ is a temperature parameter that controls the scale of the similarity score~\cite{park2020contrastive}.}
%%%%%%%%%%%%%%%%%%%%%%%%%%%%%%%%%%%%%%%%%%%%%%%%%%%%%%%%

Moreover, the effectiveness of contrastive learning depends on the \textcolor{black}{significance} of the information contained in the query~\cite{hu2022qs}. \textcolor{black}{Therefore, rather than using features from randomly selected locations, we employ the Query-selected Attention (QS-Attn)~\cite{hu2022qs} module to select queries based on their measured significance.} In our study, the QS-Attn module selects features based on the entropy of fused multi-modal features ($f^{X}_{fusion}$), thereby enabling the extraction of the most informative features from multi-source modalities. The detailed process begins by converting the $f^{X}_{fusion}$ into a 2D matrix $\mathcal{Q}$. This matrix is then multiplied by its transpose $\mathcal{K}$, resulting in a matrix that undergoes a softmax operation to generate the global attention matrix $A_g$. \textcolor{black}{For each row in $A_g$, the entropy $H_g$ is calculated using the following formula~\cite{hu2022qs}}:

\begin{equation}
 {H}_g(i)=-\sum_{j=1}^{HW}{A}_g(i,j)\log{{A}_g(i,j)}.\label{eq3}
\end{equation}
%%%%%%%%%%%%%%%%%%%%%%%%%%%%%%%%%%%%%%%%%%%%%%%%%%%%%%%%
\textcolor{black}{By measuring feature similarity, this module helps identify the most distinct feature for contrastive learning, ensuring that the network focuses on the most informative features during training~\cite{hu2022qs}.}

% Segmentation Decoder
\subsection{Segmentation Decoder}
To enhance our multi-modal translation network, we incorporate an additional segmentation decoder (${G}_{SD}$) into the framework. 
This enhancement allows the network to generate the missing target modality image and predict the segmentation output, simultaneously.
Specifically, the feature $f^{X}_{fusion}$ is fed into both the segmentation decoder (${G}_{SD}$) and the general decoder (${G}_{D}$), producing the segmentation output and the pseudo-target image, respectively.
The output from the segmentation decoder is then compared to the ground-truth segmentation maps.
The segmentation loss is calculated using the cross-entropy loss function as follows~\cite{yi2004automated}:

\begin{equation}
 {L}_{seg} = -\frac{1}{N}\sum_{i=1}^{N}\sum_{c=1}^{C}t_{i,c}\cdot{\log{(p_{i,c})}},
\end{equation}
\textcolor{black}{where $p_{i,c}$ and $t_{i,c}$ represent predicted probabilites and the ground-truth labels for class $c$ at pixel $i$, respectively~\cite{yi2004automated}.} \textcolor{black}{$N$ represents the total pixel count, and $C$ indicates the number of classes~\cite{yi2004automated}, with a total of four.}

\subsection{Overall loss function}
We employ multiple self-representation losses to ensure that target-specific information is effectively conveyed to the generator at all stages.
Similar to AE-GAN~\cite{cao2023AE}, we apply the self-representation losses within the decoder. This loss~\cite{cao2023AE} is particularly crucial and effective as it imposes a constraint ensuring the fused features from multiple source modalities remain similar to the target image features during the upsampling process.
Consequently, this self-representa-tion loss aids the multi-modal translation network in better matching the distribution of the target images, enabling the network to generate pseudo-target images that more closely resemble the target modality images. 
This self-representat-ion loss, defined as $L_{SR\_decoder}$, is calculated as the KL Divergence loss between the decoder features of ${G}_{D}$ and ${SR}_{D}$~\cite{cao2023AE}:

\begin{equation}
\mathcal{L}_{SR\_decoder} = \sum_{i=1}^{N} \text{KL}\left( \log \left(\left({G_D(f_i)} \right) \right) \parallel \left({SR_D(f_i)} \right) \right).
\end{equation}

Moreover, following ACA-GAN~\cite{cao2023ACA}, we employ two self-representation losses at different stages: $L_{SMR}$ and $L_{MMR}$. The $L_{SMR}$ is the $L2$ loss between the multi-branch encoder features and the encoder features of the self-representation network~\cite{cao2023ACA}.
The $L_{MMR}$ is the $L2$ loss between the features passing through the multi-modal fusion module and the final encoder features of the self-representation network~\cite{cao2023ACA}. 
Therefore, the total objective loss is defined as follows:
\begin{equation}
 {L}_{G}={L}_{adv}+\alpha\cdot{L}_{con}+ \beta\cdot{L}_{seg}+\gamma\cdot{L}_{SR\_decoder}+\delta\cdot{L}_{SMR}+\eta\cdot{L}_{MMR},
\end{equation}
\textcolor{black}{where $L_{adv}$ represents the adversarial loss~\cite{goodfellow2020gan}, $\alpha$, $\beta$, $\gamma$, $\delta$ and $\eta$ are hyperparameters used to balance the overall losses. $\beta$ is set to 0.05, while the others are set to 0.1.}

%% file: Chapter/3.Experiments.tex
\section{Experimental settings}
\subsection{Dataset}
The BraSyn-2023 dataset \cite{baid2021rsna,karargyris2023federated}, sourced from the RSNA-ASNR-MICCAI BraTS 2021 dataset, includes multi-parametric MRI (mpMRI) scans of brain tumors collected retrospectively across various institutions. 
The dataset comprises four modalities: pre-contrast T1-weighted (T1), post-contrast T1-weighted (T1-ce), T2-weighted (T2), and T2 Fluid Attenuated Inversion Recovery (FLAIR). 
These scans were acquired under standard clinical conditions, albeit with different equipment and protocols, leading to variations in image quality.
Expert neuroradiologists reviewed and annotated tumor subregions, including the Gd-enhancing tumor (ET), peritumoral edematous/infiltrated tissue (ED), and necrotic tumor core (NCR). 
The dataset consists of 1,251 scans for training, 219 scans for validation, and 570 scans for the private testing set.
For training, participants receive complete sets of all four modalities with segmentation labels, while in the validation and test sets, one modality is omitted to assess image synthesis methods. 
Standardized preprocessing, including DICOM to NIfTI conversion, co-registration, resampling, and skull-stripping, has been applied, ensuring high-quality data for the challenge \cite{li2023BraSyn}.
\subsection{Evaluation metric}
The inference task in the BraSyn challenge requires algorithms to predict a missing MRI modality from a test set where one of the four modalities (T1, T1-ce, T2, FLAIR) is absent.
The generated images will be evaluated on two aspects: (i) Image Quality—using the Structural Similarity Index Measure (SSIM)~\cite{wang2004image} to compare the synthesized images with real clinical images in both tumor and healthy brain regions, providing two SSIM scores per test subject; (ii) Segmentation Performance—using Dice scores for three tumor structures to evaluate the efficacy of the synthesized images in improving segmentation results. 
The segmentation will be performed using the final FeTS\footnote{https://github.com/FETS-AI/Front-End/} algorithm \cite{pati2022fets}, pre-trained on the FETS brain tumor segmentation dataset. Due to time constraints, we were unable to include segmentation results within the validation phase. However, we believe that our method has the potential to significantly enhance segmentation performance.

\subsection{Implementation Details}
We implement our proposed models using Pytorch\footnote{https://pytorch.org}.
To train the proposed model, \textcolor{black}{Adam optimizer is applied with momentum parameters $\beta_{1}$ and $\beta_{2}$ set to 0.5 and 0.999, respectively.}
The \textcolor{black}{batch size} is set to 4, and the initial learning rate is set to 0.0001. \textcolor{black}{For the training stability~\cite{cao2023ACA}, the self-representation network is pre-trained for 200 epochs. We then loaded the pre-trained parameters into the multi-modal translation network, which is trained for 300 epochs. To perform contrastive learning, we applied the global attention of QS-Attn~\cite{hu2022qs}, using an attention matrix sized $256\times256$. After sorting the attention matrix in ascending order, we utilized only the top 256 features~\cite{hu2022qs}. The features utilized for contrastive learning included those obtained after multi-modal attention as well as those processed through a convolution layer following the multi-modal attention.}
\textcolor{black}{Since our model is designed for 2D images, we transform the 1,251 3D scans into 193,905 2D slices along the axial plane. Out of these slices, only 81,437 containing tumors, as identified by segmentation masks, are used for training.}
In the inference phase, \textcolor{black}{we concatenate the 2D MRI slices generated by our model to construct the final 3D MRI imaging.}
Our four models are trained by adopting a `dedicated training' strategy that trains a series of models individually for each missing situation.
\textcolor{black}{The experimental results are analyzed for each of the four models separately on the provided validation set, using fixed missing modalities rather than randomly dropping them.}

\section{Results and Discussion}
Table \ref{T:result_official} presents the performance of our model based on the validation set from the BraSyn2024 challenge. 
Since the validation set does not contain segmentation masks, we evaluate our model on the SSIM scores for the entire MRI images, without separately analyzing tumor and non-tumor regions.
Shown in the table \ref{T:result_official}, our model achieves promising performance in generating missing FLAIR and T1-ce scenarios, achieving SSIM scores of 0.9071 and 0.9046, respectively. 
For missing T2 and T1 scenarios, the model also demonstrates exceptional performance, achieving notably higher SSIM scores of 0.9327 and 0.9258, respectively.
Moreover, the average SSIM score of 0.9182 indicates that our model maintains high image quality across various missing modality situations.

\input{table_official_val}

\begin{figure*}[hbt!]
\begin{center}
\includegraphics[width=\linewidth]{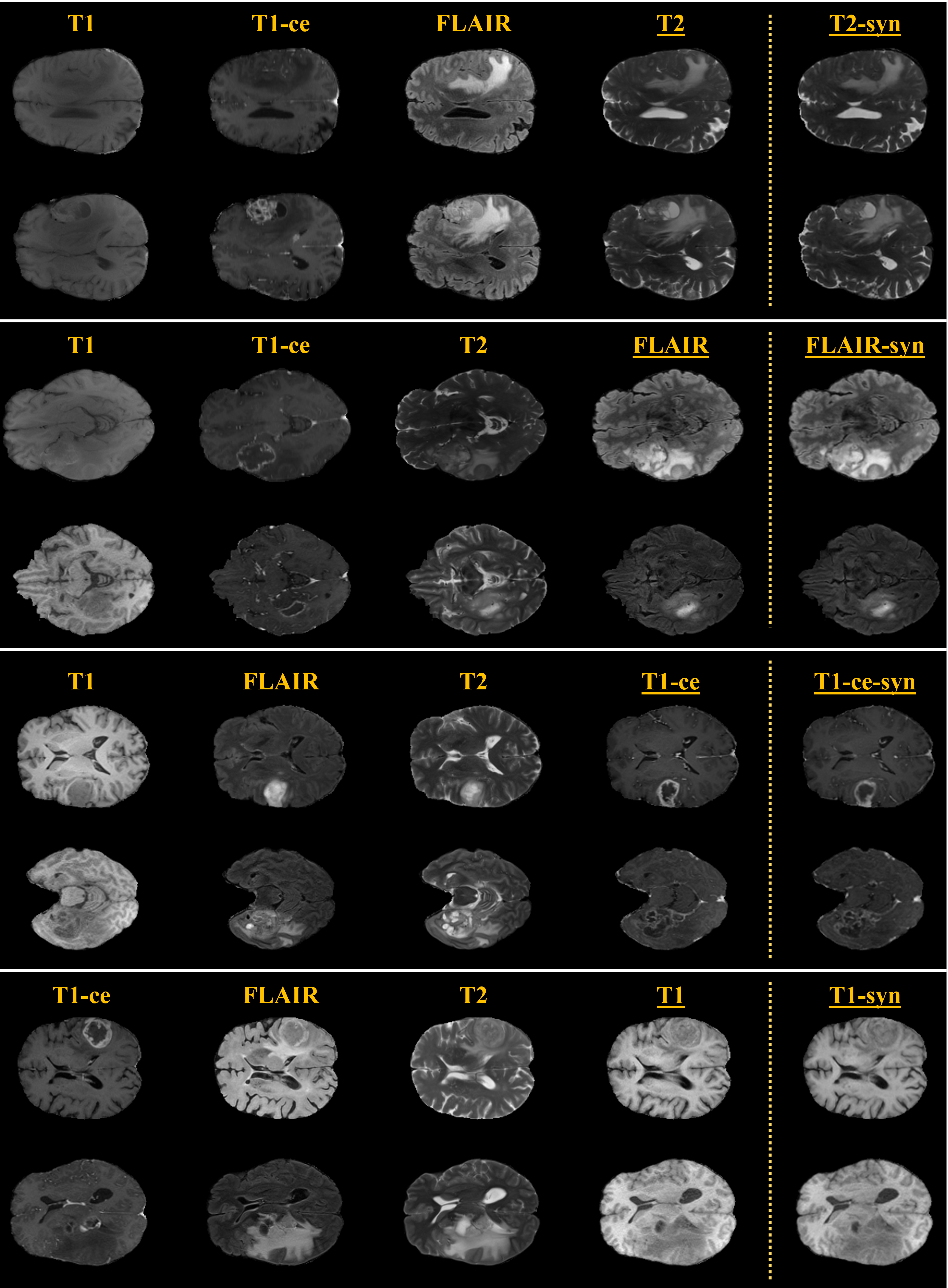}
\end{center}
\caption{Qualitative results of missing T2, FLAIR, T1-ce, and T1 scenarios. The missing target MRI generated by our model is labeled as `-syn', and the missing modality is underlined.} 
\label{fig:qualt}
\end{figure*}

Fig. \ref{fig:qualt}. shows the visual results for the qualitative evaluation of our model. For all missing scenarios, the generated target modality's soft tissue details are generally clear, and the intensity resembles the real missing modality. This indicates that our multi-modal contrastive learning approach has been effective in generating high-quality images. Moreover, the structure of tumor regions are particularly well-preserved, as our translation model contains additional segmentation decoder.

%% file: table_official_val.tex
\begin{table}[]
\caption{Quantitative results on validation dataset for BraSyn 2024 challenge. \label{T:result_official}}
\centering
\setlength{\tabcolsep}{23pt}
\renewcommand{\arraystretch}{1.2}
\resizebox{\textwidth}{!}{%
\begin{tabular}{cccc|cl}
\hline
\multicolumn{4}{c|}{\textbf{Missing situation}}             & \multicolumn{2}{c}{\multirow{2}{*}{\textbf{SSIM↑}}} \\ \cline{1-4}
\textbf{T1} & \textbf{T1-ce} & \textbf{FLAIR} & \textbf{T2} & \multicolumn{2}{c}{}                                \\ \hline
$\bullet$           & $\bullet$              & $\bullet$              & $\circ$           & \multicolumn{2}{c}{0.9327}                          \\
$\bullet$           & $\bullet$              & $\circ$              & $\bullet$           & \multicolumn{2}{c}{0.9071}                          \\
$\bullet$           & $\circ$              & $\bullet$              & $\bullet$           & \multicolumn{2}{c}{0.9046}                          \\
$\circ$           & $\bullet$              & $\bullet$              & $\bullet$           & \multicolumn{2}{c}{0.9285}                          \\ \hline
\multicolumn{4}{c|}{Average}                                & \multicolumn{2}{c}{0.9182}                          \\ \hline
\end{tabular}
}
\end{table}

%% file: Chapter/4.Conclusion.tex
\section{Conclusion}
In this work, we propose missing MRI generative
model that utilizes multi-modal contrastive learning and an additional segmentation decoder.
Our model effectively leverages crucial information from integrated multi-source modalities through the multi-modal contrastive learning.
By incorporating the segmentation decoder, we enhance the precision of tumor region generation.
Through high SSIM scores, we have demonstrated that our model generated high-quality images. Furthermore, the reported visual results of the generated images further confirms the effectiveness of our model.
Consequently, our method achieved superior performance in the BraSyn challenge, highlighting its efficacy in generating accurate and reliable target modality images.